\documentclass[twoside]{article}
\pdfoutput=1
\usepackage{qic,epsfig}
\usepackage{amsmath}

\usepackage{amssymb,color}

\newcommand{\ket}[1]{\vert #1 \rangle}
\newcommand{\bra}[1]{\langle #1 \vert}

\newcommand{\ketbra}[2]{\vert #1 \rangle \langle #2 \vert}
\begin{document}
\setlength{\textheight}{8.0truein}    %FOR 2ND PAGE ONWARDS
\runninghead{Teleportation improvement by noiseless linear 
amplification}{H. Adnane and M. G. A. Paris}
\normalsize\textlineskip
\thispagestyle{empty}
\setcounter{page}{935}
%\copyrightheading{Vol.}{No.}{Year}{Page Nos.}
%\copyrightheading{19}{11\&12}{2019}{0935--0951}
\vspace*{0.88truein}
\alphfootnote
\fpage{935}
\centerline{\bf
TELEPORTATION IMPROVEMENT BY NON-DETERMINISTIC  }
\vspace*{0.035truein}
\centerline{\bf NOISELESS LINEAR AMPLIFICATION}
\vspace*{0.37truein}
\centerline{\footnotesize HAMZA ADNANE}
\vspace*{0.015truein}
\centerline{\footnotesize\it Laboratoire de Physique Th\'eorique, Universit\'e de Bejaia, Campus Targa Ouzemour, 06000 Bejaia, Algeria}
\vspace*{10pt}
\centerline{\footnotesize MATTEO G. A. PARIS}
\vspace*{0.015truein}
\centerline{\footnotesize\it Quantum Technology Lab, Dipartimento di Fisica {\em Aldo Pontremoli}, Universit\`a degli 
Studi di Milano, I-20133 Milano, Italy}
%%
%\publisher{May 14, 2019}{August 9, 2019}	
\vspace*{0.11truein}
\abstracts{We address de-Gaussification of continuous 
variables Gaussian states by optimal non-deterministic noiseless 
linear amplifier (NLA) and analyze in details the properties of 
the amplified states. In particular, we investigate the entanglement 
content and the non-Gaussian character for the class of non-Gaussian 
entangled state obtained by using NL-amplification of two-mode squeezed 
vacua (twin-beam, TWB). We show that entanglement always increases, whereas 
improved EPR correlations are observed only when the input TWB has low energy.
We then examine a Braunstein-Kimble-like protocol for the teleportation 
of coherent states, and compare the performances of TWB-based teleprotation 
with those obtained using NL-amplified resources. We show that teleportation
fidelity and security may be improved for a large range of NLA parameters 
(gain and threshold).}{}{}	
\vspace*{10pt}
%%%	
\vspace*{3pt}
%\communicate{S~Braunstein~\&~J~Eisert }
\vspace*{1pt}\textlineskip
%%%
\section{Introduction}
\noindent
Quantum teleportation has proven to be one of the most relevant manifestations 
of entanglement~\cite{Nielsen,Bennett}. In this protocol, an unknown quantum state is 
teleported from a sending station to a remote receiving terminal by 
exploiting a quantum channel made by a two mode entangled state. In the framework 
of continuous-variable (CV) systems several teleportation protocols exploiting Gaussian entangled optical resources have been suggested and experimentally 
realized~\cite{Braunstein,Furusawa,Yukawa}. The Gaussian entangled resource 
commonly employed in CV quantum information protocols is the two-mode 
squeezed vacuum state (also termed twin-beam or EPR state) that may be produced 
by the process of parametric downconversion. Nonetheless, it has been shown that Gaussian entangled resources as twin-beams and any bipartite Gaussian states 
generated by Gaussian operations are subjected to restrictions. For instance, 
local Gaussian operations applied to bipartite Gaussian states cannot lead to entanglement distillation~\cite{Eisert, Fiurasek}. Likewise, in the framework of quantum communication protocols, it has been shown that the security of Gaussian states against Gaussian errors cannot be improved by Gaussian operations~\cite{Niset}. Besides those restrictions, engineering high squeezed twin-beams is an arduous task even when the down conversion process is implemented in a resonant cavity. In fact, the highest amount of squeezing for the twin-beam attainable with current technology is around 10 dB~\cite{Eberle} whereas the state-of-the-art of single mode squeezing boarder on 15 dB~\cite{Vahlbruch}. 
\par
Thereby, a particular attention was drawn to non-Gaussian entangled resource that are more suited for quantum communication tasks. For this purpose, divers de-Gaussification protocols have been considered previously so as to improve the efficiency of quantum information processing. In particular, photons subtraction, photons addition and their coherent superposition performed on twin-beams has proven to enhance the degree of entanglement along with the teleportation fidelity~\cite{Opatrny, Cochrane, Dell'Anno, Yang, Olivares, Olivares2}.
Recently, our investigations~\cite{Adnane} on the action of an optimal non-deterministic noiseless linear amplifier (NLA)~\cite{Menzies, McMahon, Pandey} on twin-beams revealed its robustness for the generation of non-Gaussian bipartite states and the enhancement of the degree of entanglement captured by the excess Von Neumann entropy~\cite{Barnett, Vedral}. Previous works have demonstrated that some particular properties of non-Gaussian entangled resources are crucial to enhance the quality of quantum teleportation. Particularly, Dell'Anno \textit{et. al}.~\cite{Dell'Anno} showed that optimization of the teleportation with a B-K-like protocol using certain non-Gaussian resources is made possible by adjusting their amount of entanglement, non-Gaussianity (NG) and squeezed vacuum affinity (SVA). Moreover, in~\cite{Braunstein}, Einstein-Podolsky-Rosen (EPR) correlation of the entangled resources is pointed out to be an indicator for quantum teleportation.
\par
In this work, the characteristics of the successfully amplified twin-beams generated from the action of the optimal NLA along with their performance on teleportation of coherent states are investigated. The degree of entanglement quantified by the excess Von Neumann entropy, the non-Gaussianity and the EPR correlations of the non-Gaussian twin-beams are discussed. Their performances in (CV) quantum teleportation of coherent states in a  B-K-like protocol are then assessed. Finally, we identify the characteristics of the non-Gaussian amplified twin-beams that lead to an improvement of the teleportation success (quantified by the average fidelity of the output state with respect to the input unknown one), compared to the standard twin-beams. We will show that substantial improvements of the teleportation success are achieved in the regime where the EPR correlation of the entangled resource are lowered while a non-trivial dependence on the degree of entanglement and the non-Gaussianity quantified by the quantum relative entropy~\cite{Genoni}, is reported. It's worth noting that noiseless amplification was previously proposed to be employed in a teleportation set-up to achieve continuous variable error correction on Gaussian states that experienced Gausssian noise occasioned by loss~\cite{Ralph2}. 

\bigskip
The paper is structured as follows: in Sec~\ref{s:sec2}, we analyse relevant 
properties of two-mode non-Gaussian entangled states engineered via an 
optimal NLA and confront them with that of the standard twin-beam. 
Entanglement content captured by the excess Von Neumann entropy and
EPR correlation along with the entropic non-Gaussianity are discussed. 
Sec~\ref{s:sec3}, is devoted to continuous-variables quantum teleportation 
with a class of non-deterministically amplified twin-beams. The paradigmatic 
instance of coherent states as inputs is discussed. Finally, in Section
\ref{s:sec4} we summarize our results and set forth our conclusions.
\section{Characteristics of the amplified twin-beams}\label{s:sec2}
\setcounter{footnote}{0}
\noindent
Aiming to examine the properties of the non-Gaussian amplified twin-beam, we first review some details regarding its generation and recall certain characteristics of the Gaussian standard twin-beam.
\subsection{The standard twin-beam bipartite state}
\noindent
The two-mode squeezed vacuum is a broadly used entangled resource in divers quantum information protocols. Its generation involves the process of parametric down-conversion where a $\chi^{(2)}$ non-linear crystal serving as an amplifying medium is pumped with light at frequency $\omega_{p}$. A fraction of it gives then arise to a pair of photons with frequencies $\omega_{a}$ and $\omega_{b}$, obeying to the constraint $\omega_{p}=\omega_{a}+\omega_{b}$.
Formally, its expression is obtained by applying the two mode squeezing operator $S_{ab}(r)=\exp{-r(\hat{a}^{\dagger}\hat{b}^{{\dagger}}-\hat{a}\hat{b})}$~\cite{Loudon} to the vacuum:
\begin{equation}
\left\vert \chi\right\rangle =\exp{(r(\hat{a}^{\dagger}\hat{b}^{{\dagger}}-\hat{a}\hat{b}))}\left\vert 00 \right\rangle =\sqrt{1-\chi^{2}}%
{\displaystyle\sum\limits_{n=0}^{\infty}}\chi^{n}\left\vert nn\right\rangle , \label{1}%
\end{equation}
where $r$ is assumed real without loss of generality,  $\hat{a}^{\dagger}$($\hat{a}$) and $\hat{b}^{\dagger}(\hat{b})$ are the creation(annihilation) operators of the two modes and $0<\chi=\tanh{r}<1$ depends on the non-linear susceptibility and an effective interaction length.
The standard twin-beam has a Gaussian Wigner function. Since it is a pure (CV) state, a good measure of its entanglement content is the excess Von Neumann entropy defined as the Von Neumann entropy of the reduced density operator $E\left[\varrho_{ab}\right]  =\hat{S}\left[  \rho_{a}\right]  =-Tr\left[\rho_{a}\ln\rho_{a}\right]$~\cite{Araki}, where the reduced density operator $\rho_{a}$ is obtained by tracing over the mode $b$ of the twin-beam. Its Von Neumann entropy is found to be 
\begin{equation}\label{2}
S\left[  \rho_{a}\right]=-\ln{(1-\chi^{2})}-\frac{ \chi^{2}\ln{(\chi^{2})}}{(1-\chi^{2})}.
\end{equation}

Besides the degree of entanglement characterizing an entangled resource, Einstein-Podolsky-Rosen correlation of a bipartite state has proven to be a prominent element to carry out quantum teleportation. Instead of the position and momentum operators of a massive particle discussed in the celebrated paper on the completeness of quantum mechanics~\cite{Einstein}, the quantities of interest for an optical system are the quadrature operators defined as

\begin{equation} 
\begin{split}
 \hat{x}_{a} =\frac{1}{\sqrt{2}}(\hat{a}+\hat{a}^{\dagger}),\qquad  \hat{p}_{a}=\frac{-i}{\sqrt{2}}(\hat{a}-\hat{a}^{\dagger}),
\\
\hat{x}_{b}=\frac{1}{\sqrt{2}}(\hat{b}+\hat{b}^{\dagger}) ,\qquad  \hat{p}_{b}=\frac{-i}{\sqrt{2}}(\hat{b}-\hat{b}^{\dagger}),
\end{split}
\end{equation}
Regarding a bipartite optical state, the EPR correlation reads ~\cite{Duan}
\begin{equation}
\Delta z^{2}=\Delta(\hat{x}_{a}-\hat{x}_{b})^{2} +\Delta(\hat{p}_{a}+\hat{p}_{b})^{2},
\end{equation}
which, after expressing the quadrature operators in term of the annihilation and creation operators becomes
\begin{equation}\label{EPR}
\Delta z^{2}=2\left[ 1+\left \langle \hat{a}^{\dagger}\hat{a}\right \rangle + \left \langle \hat{b}^{\dagger}\hat{b}\right \rangle -\left \langle \hat{a}^{\dagger}\hat{b}^{\dagger} \right \rangle - \left \langle \hat{a}\hat{b} \right \rangle -(\left \langle \hat{a} \right \rangle-\left \langle \hat{b}^{\dagger} \right \rangle)(\left \langle \hat{a}^{\dagger} \right \rangle-\left \langle \hat{b} \right \rangle)\right] .
\end{equation}

A zero EPR correlation reveals perfect correlation between the two modes as for the ideal EPR state introduced in~\cite{Einstein}. Indeed, both variances of the position-like operators difference and momentum-like operators sum are zero, thus indicating that knowing the quadratures of say mode '$a$' enables to exactly determine the quadratures of the mode '$b$'. 
Contrariwise, a value exceeding 2 indicates classical separable two-mode states. For a two-mode squeezed vacuum, the EPR correlation is found to be

\begin{equation}
\Delta z^{2}=\frac{2(1-\chi)}{(1+\chi)}.
\end{equation}
As long as the squeezing parameter $\chi$ belongs to $]0,1[$, the EPR correlation of the standard twin-beam remains lower than 2, thus indicating the presence of quantum correlations. Moreover, when $\chi$ broads on 1 (in the limit of infinite squeezing), the EPR correlation becomes null, testifying for maximum entanglement while in the case of weak squeezing, it approaches 2. Thereby, for a two mode squeezed state, highest squeezing strength lead to strongest correlations. Finally, we recall that at fixed energy, the twin-beams are known to be the maximally entangled (CV) bipartite states~\cite{Popescu}.

\subsection{The amplified twin-beam}\label{amplifiedTB}
Before going deeper into our survey, we propose to expose in short, some of the most emblematic experimental realisations of the non-Gaussian operation discussed here, namely, the noiseless linear amplification. Since the pioneer work of Ralph and Lund~\cite{Ralph1} where the idea of probabilistic noise-free amplifiers was coined, several implementations of the NLA has been achieved by different groups. The first realisations were physical: a suitably engineered device involving optical components was required. The operating principles of the proposed schemes range from generalized quantum scissors involving auxiliary single-photon sources and multiphoton interferometric set-ups~\cite{Xiang,Ferreyrol,Ferreyrol1} to coherent superposition of single photons addition and subtraction~\cite{Jarumir,Zavatta}, and include concentration of phase information via noise addition~\cite{Usaga}. Thereafter, given the demanding resources involved in the physical designs, virtual noiseless amplifiers were theoretically examined~\cite{Jarumir1,Walk} then experimentally achieved~\cite{Chrzanowski,Haw}. Surprisingly, these alternative schemes, commonly termed "Measurement-based NLA" only require a postselection of heterodyne detection outcomes that emulate the quantum filter induced by the physical non-deterministic amplification. In the end, each of the designs mentioned so far along with the optimal architecture considered in our work~\cite{Pandey} naturally afford different performances in terms of success probability and fidelity to the ideally amplified input, and are found to be approximate realizations of the ideal noise-free phase-insensitive amplifier achieving the operation $\hat{g}^{a^{\dagger}a}$.
\par
Generation of non-Gaussian entangled bipartite state via
photon subtraction(addition) or their coherent superposition
on two-mode squeezed vacuum states was revealed useful
in quantum information processing. Likewise, non-Gaussian
states produced by mixing photon subtracted squeezed vacuum
states with vacuum in a beam splitter showed interesting
improvement in the quality of quantum teleportation
of coherent states~\cite{Bose}. Here we discuss a process of 
de-Gaussification based on the action of an optimal NLA on
a twin-beam. One mode of the twin-beam is coupled to a
measurement device (MD) consisting of a two-level system (as it happens, $\left| S \right\rangle  $ and  $\left| F \right\rangle $, where "$S$" refers to success whereas "$F$" stands for failure)
through a unitary transformation. The (MD) is then projected
into one of its initial states $\left| S \right\rangle  $ or  $\left| F \right\rangle $, that heralds the success
or the failure of the amplification. We consider only the
successfully amplified states that we refer to as the ”amplified
twin-beam”. The process is thus inherently conditional and
was presented in details in~\cite{Adnane}. Here we just recall the expression
of the amplified twin-beam derived from the action of the Krauss operator accounting for a successful run of the optimal device~\cite{Pandey,McMahon}
\begin{align}
\label{TWB}
\left\vert \chi_{s}\right\rangle = & \frac{\hat{E}_{s}^{p}\otimes\mathbb{I}_b\left\vert
	\chi\right\rangle}{\sqrt{P_{s,\chi}}} \\
=&\sqrt{\frac{{1-\chi^{2}}}{{P_{s,\chi}}}}\left(  g^{-p}{\displaystyle\sum\limits_{n=0}^{p}}(g\chi)^{n}\left\vert nn\right\rangle+
{\displaystyle\sum\limits_{n=p+1}^{\infty}}\chi^{n}\left\vert 
nn\right\rangle \right) , 
\end{align}
where $g$ and $p$ are the intrinsic parameters of the optimal NLA
and denote respectively the gain and the integer setting the
truncation order in the Fock basis that we dub ”threshold”.
$P_{s,\chi} $ represents the probability to successfully implement the desired
amplification on an initial twin-beam and reads
\begin{equation} \label{Pschi}
P_{s,\chi}=(1-\chi^{2})\left[  g^{-2p}
{\displaystyle\sum\limits_{n=0}^{p}}
(g\chi)^{2n}+\sum_{n=p+1}^{\infty}\chi^{2n}\right].
\end{equation}

\begin{figure} [htbp]
	%\vspace*{13pt}
	\centerline{\epsfig{file=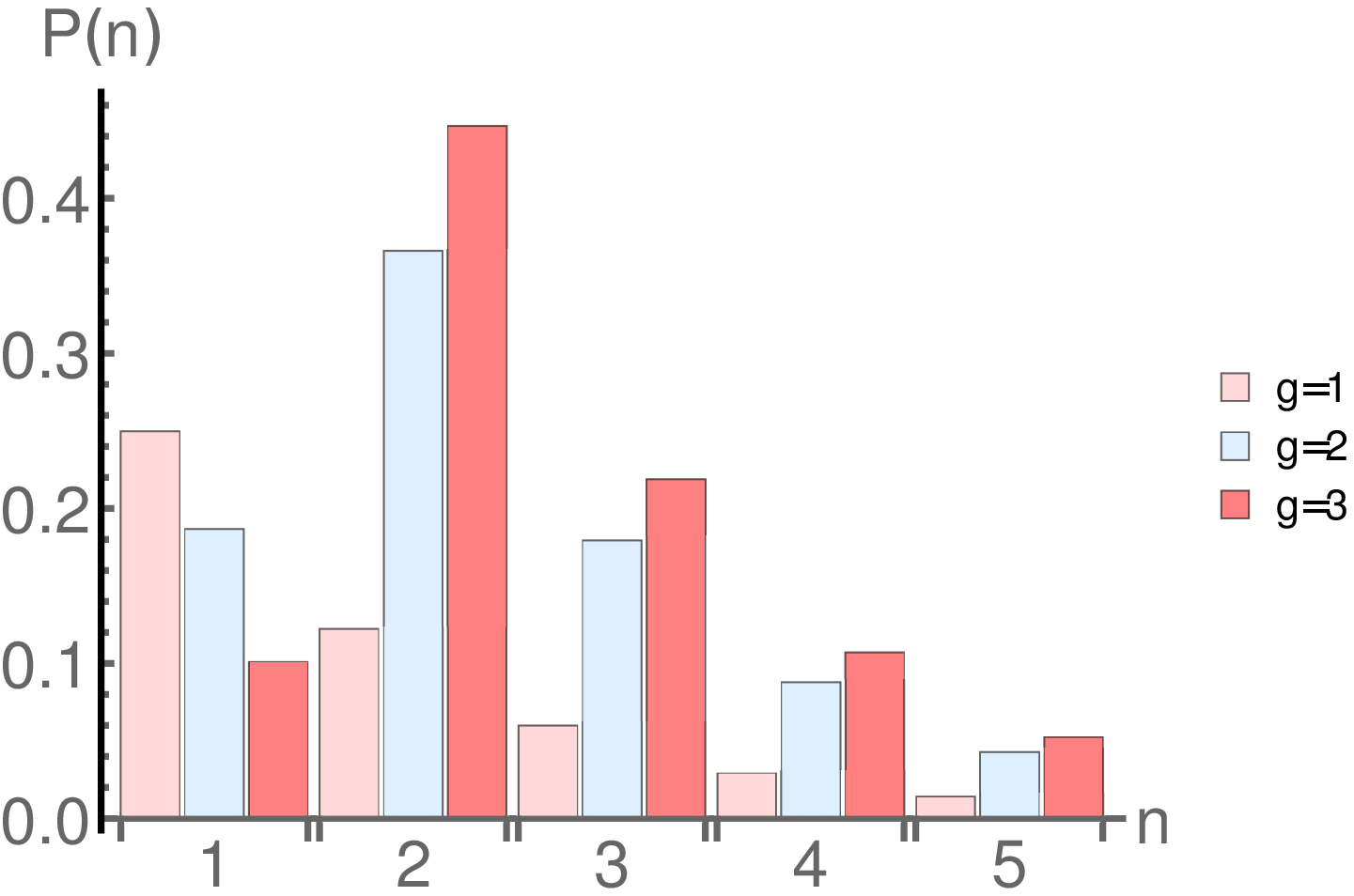, width=8.2cm}} %100 percent
	\vspace*{13pt}
	\fcaption{(Color online) Bar chart representation of the photon number distribution for various bipartite states: the standard twin-beam $g=1$ (light-red bar) and amplified twin-beams generated through different configurations: $g=2$ (light-blue bar), $g=3$ (pink bar). The threshold is kept fix $p=2$ and the squeezing parameter is setted equal to $0.6$.}\label{fig:photdist}
\end{figure}

As we clearly see, the NLA tends to privilege the occurrence
of higher photon numbers, thus inducing a modification
in the photon number distribution. Fig.~\ref{fig:photdist} shows the
bar chart representation of the photon number distribution assigned to the amplified
twin-beams and their standard version for a fixed squeezing
parameter and different configurations of the NLA. We notice
that indeed, the coincidence of measuring higher photons
number in the two modes of the amplified twin-beam has a
greater weighting then its standard version while the inverse
is observed for single photons. Moreover, we remark that
stronger is the amplification (great values of the gain), more
pronounced is the variation of the photon number distribution
induced by the NLA. This observation remains valid
for any values of the squeezing parameter $\chi$ and the threshold
$p$.
\par
We remind the reader that earlier works~\cite{Ralph1,McMahon} have considered in some details the effect of a noiseless linear amplifier on EPR states. In fact, in the pioneer survey by Ralph and Lund~\cite{Ralph1}, the authors examined the action of a particular NLA consisting of multiple blocks of quantum scissors on EPR states both for ideal and lossy channels. On the other hand, a different theoretical description of the amplifier has been examined in\cite{McMahon}, where performances of an optimal device on a lossy EPR state are drawn up.
\subsection{Non-Gaussianity}

As emphasized previously, non-Gaussianity (nG) has proven to be a resource for various tasks in quantum information
processing. In order to examine the effect of non-Gaussianity on
the performances of the amplified twin-beam, we make use of the 
entropic ”measure” introduced in~\cite{Genoni}.
For a generic quantum state , the entropic nG measure is defined
as the quantum relative entropy between the state under
study and its reference Gaussian state $\hat{\varrho_{G}}$  (that is completely
defined through its vector of mean values and covariance matrix)
\begin{equation}\label{nG}
\delta_\mathrm{nG}[\hat{\varrho}]=S(\hat{\varrho}_G)-S(\hat{\varrho}) \, ,
\end{equation}
where $S(.)$ remains for the Von Neumann entropy. The amplified
twin-beam being pure, its Von Neumann entropy vanishes
and the evaluation of its non-Gaussian character reduces to the
Von Neumann entropy of  $\hat{\varrho_{G}}$. For a generic bipartite Gaussian
state, the Von Neumann entropy is given by
\begin{equation}
S(\hat{\varrho})=h(d_{+})+h(d_{-}) \, ,
\end{equation}
where $d_{+}$ and $d_{-}$ are the symplectic eigenvalues of the covariance
matrix assigned to the amplified twin-beam and $h(x)$ a function
defined as
\begin{equation}
h(x)=\left (x+\frac{1}{2}\right )\ln\left (x+\frac{1}{2}\right )
-\left( x-\frac{1}{2}\right )\ln\left ( x-\frac{1}{2}\right ).
\label{7}
\end{equation}
The entropic nG measure is then found to be
\begin{equation}
\delta_\mathrm{nG}\Big[\ketbra{\chi_s}{\chi_s}\Big]=2 h(d_+)=2 h(\sqrt{I_1+I_3}) \, 
\end{equation}
where $I_{1}=\frac12 +\bar{N}_\chi$ and $I_{3}=\text{Tr}[\ketbra{\chi}{\chi}(\hat{a}\hat{b})]$. $\bar{N}_\chi$ 
denotes the average photons number in one mode of the amplified twin-beam. For 
detailed calculations, see~\cite{Adnane}. 
\begin{figure} [htbp]
	%\vspace*{13pt}
	\centerline{\epsfig{file=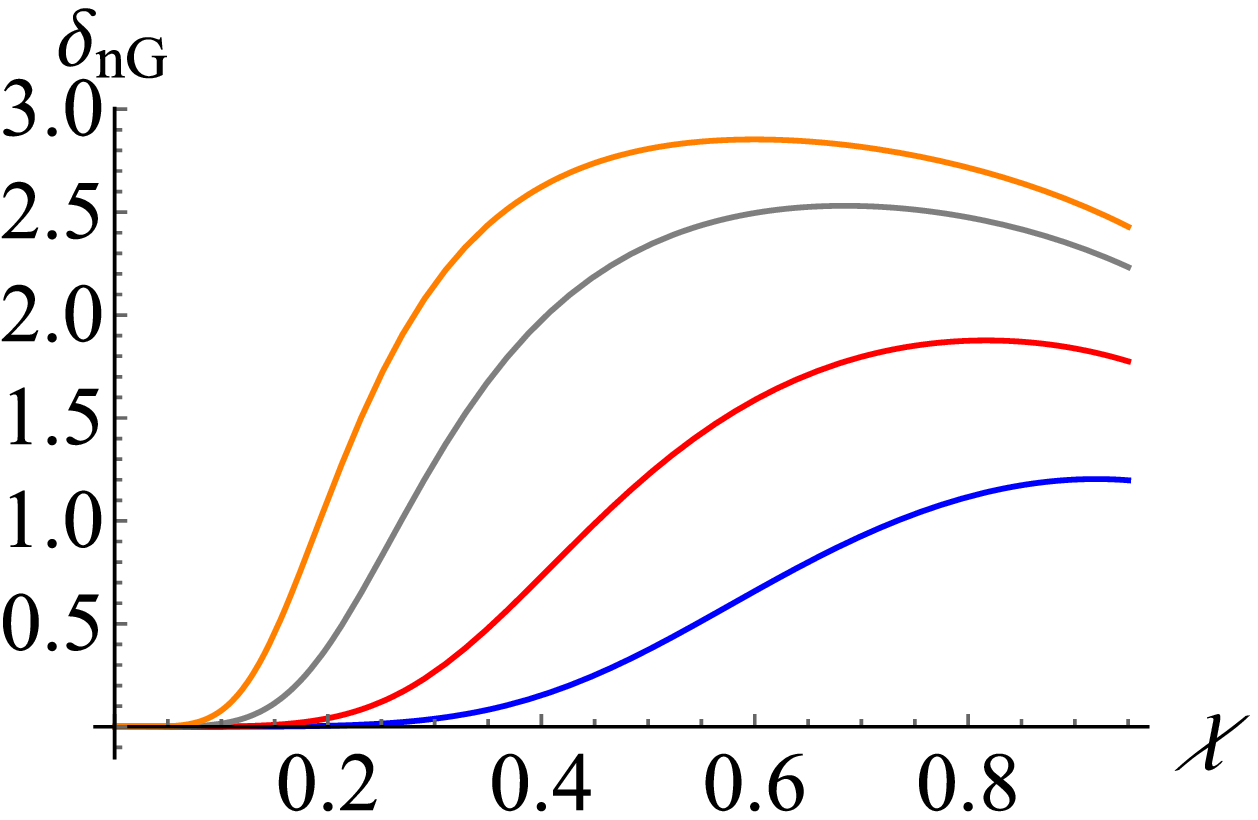, width=8.2cm}} %100 percent
	\vspace*{13pt}
	\fcaption{(Color online) 
		Entropic Non-Gaussianity measure $\delta_{nG}$ as a function of the
		squeezing parameter for the bipartite states generated by an optimal
		NLA with different calibrations. The threshold p is fixed to 2 while
		varying the gain: the blue dotted line represents $g=1.5$ , the red
		dashed line $g=2$, the grey dotted-dashed line denotes $g=3$ and
		finally, the orange full line $g=4$.}\label{fig:non-Gaussianity}
\end{figure}
In Fig.~\ref{fig:non-Gaussianity}
are reported the plots of the entropic non-Gaussianity of the
amplified twin-beam as a function of the squeezing parameter.
Different configurations of the NLA are considered. We
observe a non-monotonic behaviour for the entropic nG with
respect to the squeezing parameter $\chi$. Two regimes are clearly noticeable: first, the nG increases with the squeezing parameter
to reach its peak then tends to diminish softly. The rate at
which the entropic nG increases, the value at which the peak is
achieved and its magnitude depend on the configuration of the
NLA. We observe that the more intense is the configuration, the swiftest is the increment of the nG and the higher is its peak.
Moreover, for intense configurations, the peak is reached at
lower squeezing parameters, thus indicating that starting from
a weakly squeezed twin-beam, one could engineer highly 
non-Gaussian bipartite state via the NLA. Finally, we notice that
for relatively weak amplifications and squeezing parameters,
the emerging bipartite state remains Gaussian.
\subsection{Degree of entanglement}

\begin{figure} [htbp]
	%\vspace*{13pt}
	\centerline{ \epsfig{file=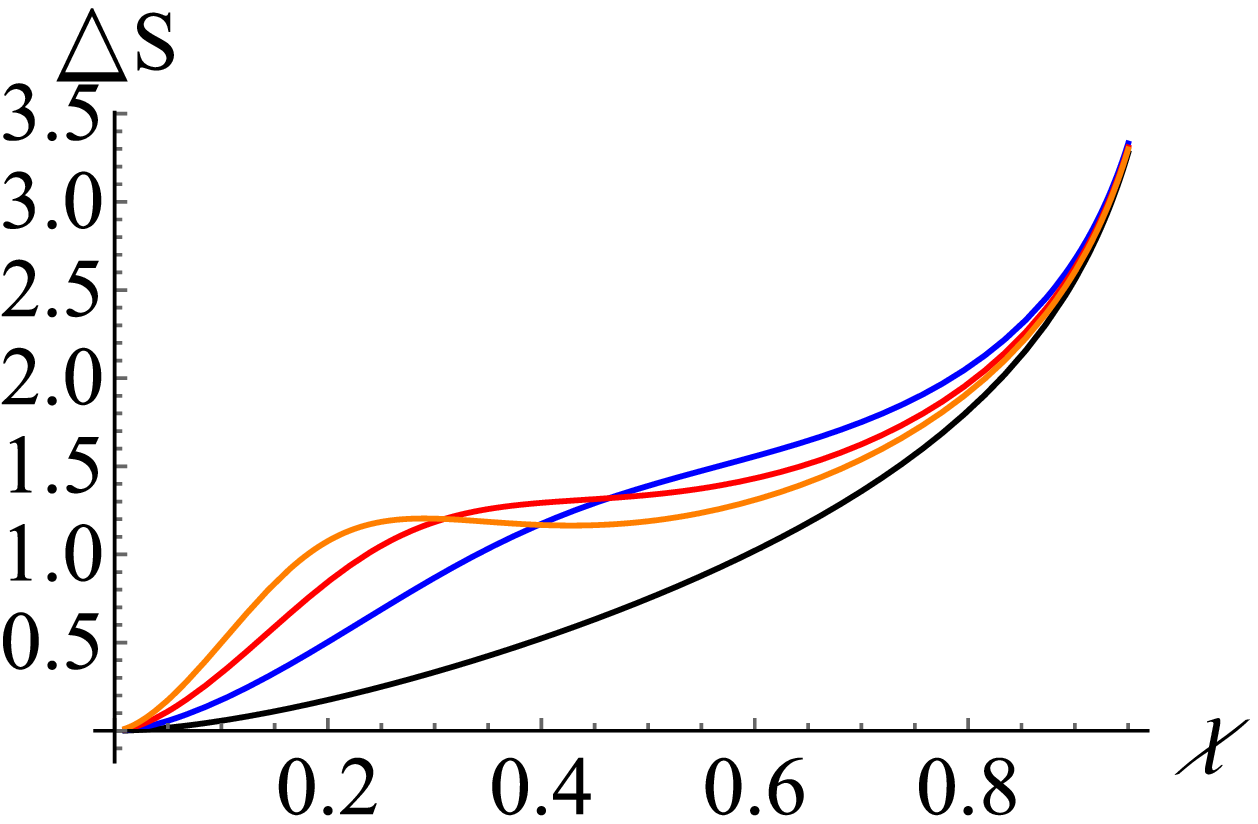, width=6.5cm}
		\epsfig{file=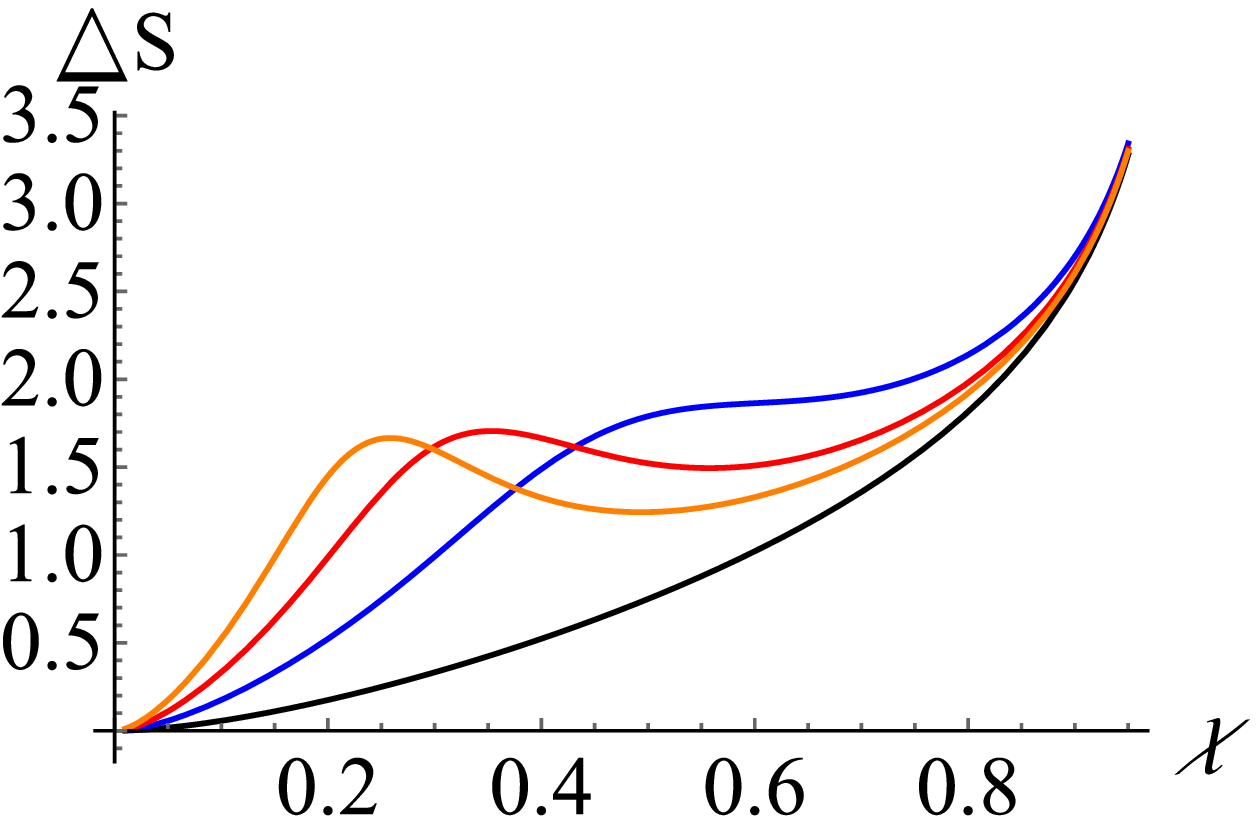, width=6.5cm}}
	
	\vspace*{13pt}
	\fcaption{(Color online) Excess Von Neumann entropy of the amplified
		and standard twin-beams as a function of the squeezing parameter.
		Different NLA configurations are considered: the blue dotted line
		denotes $g=2$ , the red dashed line $g=3$ and the orange dotted-dashed
		line represents $g=4$. The black full line remains for the
		standard twin-beam. In the left panel, $p=2$ whereas in the right
		panel $p=4$.} \label{f:ExcessVN}
\end{figure}

Here we intend to quantify the effect of the optimal NLA on
the entanglement content of the twin-beam. From the moment
that the resulting amplified twin-beam remains pure (the quantum
operation describing a successful amplification is represented
by a single Krauss operator~\cite{Pandey}) we turn again towards the
Von Neumann excess entropy to quantify its degree
of entanglement. In Fig.~\ref{f:ExcessVN} are reported the plots of the excess
Von Neumann entropy as a function of the squeezing parameter
for different configurations of the NLA. The threshold is
kept fixed while varying the gain. For the sake of comparison,
we also report the entanglement content of the standard twin-beam.
As expected, all the considered quantities are increasing
functions of the squeezing parameter. We remark that the
degree of entanglement of the amplified twin-beams is greater
than that of the standard bipartite squeezed vacuum, and that
for all the considered configurations and at any value of the energy,
particularly in the interval $0 < \chi < 0,75$ experimentally
available. Moreover, strong amplifications tend to enhance the
degree of entanglement of twin-beams characterized by weak
energies while the opposite is observed for weak amplification.
Finally, as observed for the entropic non-Gaussianity, the
more intense are the NLA configurations, lowest are values of
the squeezing parameter at which the peaks of the excess Von
Neumann entropy are reached.
\subsection{EPR correlation}

\begin{figure} [htbp]
	%\vspace*{13pt}
	\centerline{ \epsfig{file=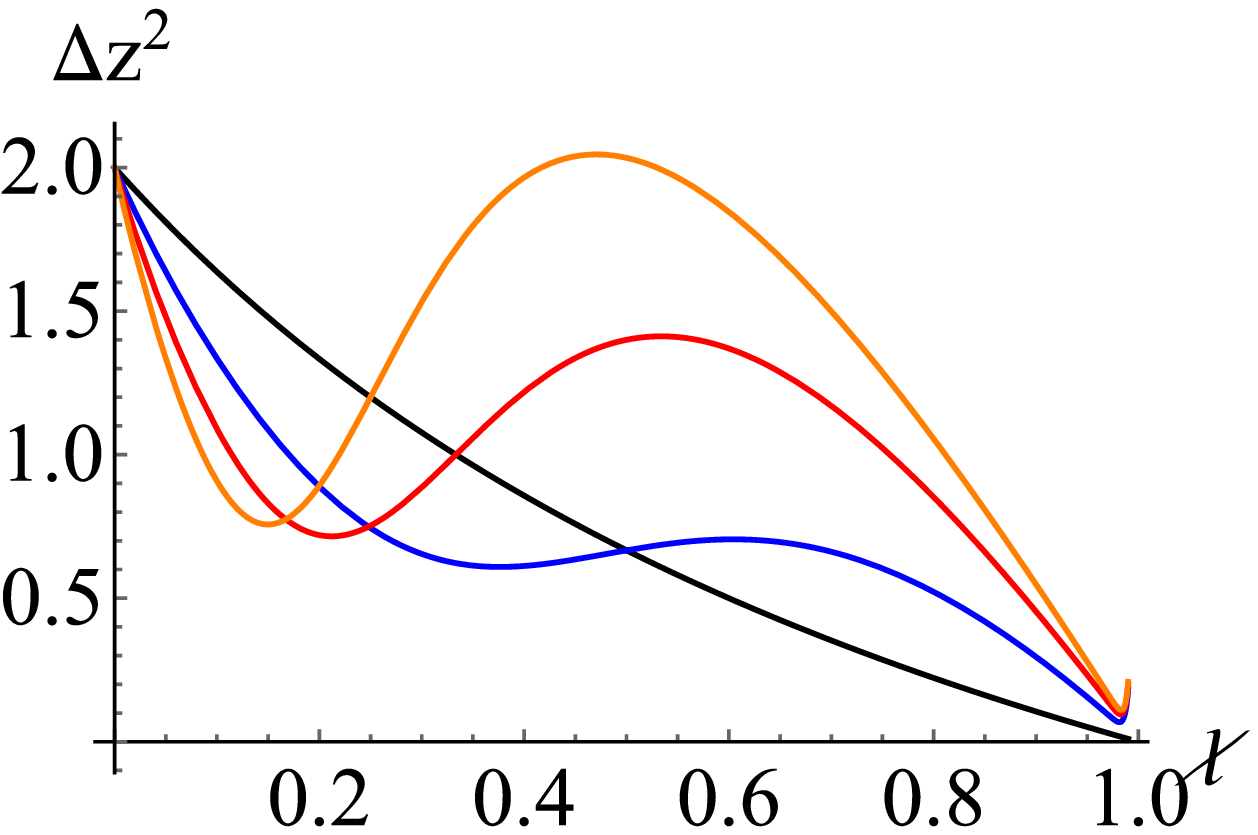, width=6.5cm}
		\epsfig{file=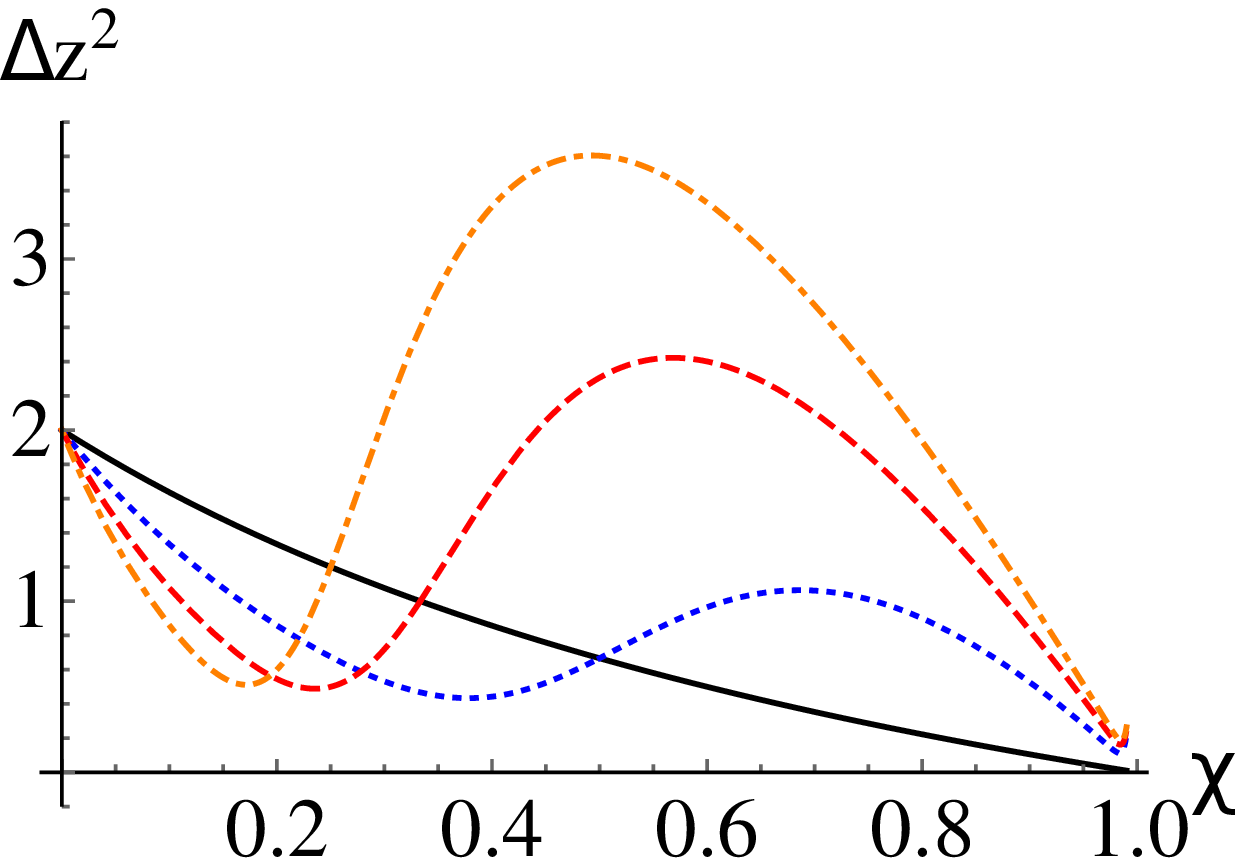, width=6.5cm}}
	
	\vspace*{13pt}
	\fcaption{(Color online) The EPR correlation $\Delta z^{2}$ as a function
		of $\chi$ for a two mode squeezed vacuum (black full line) and amplified
		twin-beams generated from different configurations of the NLA:
		$g=2$ (blue dotted line), $g=3$ (red dashed line), $g=4$ (orange
		dotted-dashed line). In the left panel, $p$ is setted to 2 while in the
		right panel $p=4$.} \label{f:EPR}
\end{figure}

Previously, we addressed EPR correlation between the optical
field quadratures as an alternative approach to quantify
entanglement content of a bipartite state. In This subsection,
we discuss the effect of an optimal non-deterministic
noiseless amplification on the EPR correlation of a two-mode
squeezed vacuum. We underline that for the bipartite states
of interest, the variance of the position difference and the momentum
sum are equal, thus the EPR correlation resumes to
$\Delta z^{2}=2\Delta\left(\hat{x}_{a}-\hat{x}_{b}\right)^{2}$. 
Fig.~\ref{f:EPR} shows the variations of
the EPR correlation with respect to the squeezing parameter.
Different configurations where the threshold is fixed and the
gain varies are considered and compared to that of the standard
twin-beam. We remark that apart from the standard twin-beam
whose position difference variance is a monotonically
decreasing function of the squeezing parameter, the quantity
of interest shows a non-monotonic behaviour for all the generated
bipartite states from different configurations. We also
notice that if $\chi$ is smaller than a threshold that depends on the
intrinsic parameters of the NLA, the EPR correlation is lower
than that of the standard twin-beam. Thereby witnessing for
the robustness of an optimal non-deterministic noiseless amplifier
acting in a certain regime to increase the entanglement
content of a twin-beam of a weak energy. In the contrary, beyond
that threshold, the variance of the amplified twin-beams
is larger than that of the standard two-mode squeezed vacuum.
Especially, regarding strong amplifications, the EPR correlation
attains values exceeding 2 that separate quantum entangled
bipartite states from separable states, thus testifying
for the inefficiency of the NLA in that range of energies.

According to Fig.\ref{f:ExcessVN} and Fig.\ref{f:EPR}, the NLA enhances the entanglement content captured by the excess Von Neumann entropy
of a twin-beam regardless of its initial squeezing parameter
while a different effect is observed when entanglement is
quantified by the EPR correlation, where enhancement is observed
only in the interval of weak energies. We conclude
that an increase of the excess Von Neumann entropy doesn’t
systematically imply strongest EPR correlation.
In the light of our upshot, one may expect the non-Gaussian
amplified twin-beams to enhance the fidelity of quantum teleportation
in the range of weak squeezing parameter. This
stems from the fact that in one hand, the Braunstein-Kimble
protocol is based on two-mode squeezed vacua as entangled
resources that are subjected to drastic restrictions~\cite{Eisert}, in the
other hand, we showed that the entanglement content of the
non-deterministically amplified twin-beams captured by both
the excess Von Neumann entropy and the EPR correlation is
improved by the action of the NLA in the range of weak energies.
We recall that the interval extent depends on the NLA
intrinsic parameters.
\section{Teleportation improvement assisted by NLA}\label{s:sec3}
Quantum teleportation was initially proposed by Bennet \textit{et.
	al~.} in the discrete variable regime where an unknown quantum
state of a half-spin particle is transferred from a sending
station to a distant receiver via a dual combination of a quantum
channel consisting of EPR states and classical communication.
Later Vaidman proposed a similar scheme in the
domain of continuous-variable (CV) systems where the information
is encoded in the position and momentum of the system~\cite{Vaidman2}
. Thereafter, an example of (CV) quantum teleportation
in the optical domain was presented by Braunstein and
Kimble where information is encoded in the quadratures of
the optical system and the two-mode squeezed vacuum used
as an entangled resource~\cite{Braunstein}. Subsequently, Furusawa and collaborators
implemented the protocol experimentally~\cite{Furusawa}. Here we
present our theoretical results on (CV) quantum teleportation assisted by
NLA where the entangled resource is a non-deterministically
amplified two-mode squeezed vacuum. We further draw up a
comparative study with respect to the conventional B-K protocol
with the standard twin-beam. We are particularly interested
in the teleportation of coherent Gaussian states. Foremost,
let us swiftly recall the main steps of the standard BK
protocol. Two protagonists Alice and Bob, share beforehand
an entangled bipartite resource of modes $a$ and $b$. Alice
possess a single-mode input unknown state $\left|\psi_{in}\right\rangle $ that she intends
to transfer to Bob and thus corresponds to the state to
be teleported. Her task is achieved as follows: she mixes the
state in mode ”in” on her possession with the mode $a$ of the
shared bipartite resource state at a balanced beam splitter, resulting
in the output modes $c$ and $d$. An homodyne detection
is then performed by Alice on the output modes and the result $\beta=x_{-}+ip_{+}$ transmitted to Bob through a classical channel.
The beam splitter being balanced, the real and imaginary
parts of $\beta$ are respectively the eigenvalues of the position-like
difference and momentum-like sum
\begin{equation}\label{quad} 
\frac{1}{\sqrt{2}}(\hat{x}_{in}-\hat{x}_{a}),\qquad  \frac{-i}{\sqrt{2}}(\hat{p}_{in}+\hat{p}_{a}),
\end{equation}
where $\hat{x}_{in}$, $\hat{p}_{in}$ are the quadrature operators of the mode ”in”
and $\hat{x}_{a}$, $\hat{p}_{a}$  those of the mode ”a”. The quadrature operators
appearing in Eq.~(\ref{quad}) are the observables measured through
the homodyne detection and are related to the input modes
operators via the unitary transformation induced by the $50/50$
beam-splitter. Finally, Bob applies a displacement $\hat{D}(\beta)=\exp{\left(\beta\hat{b}^{\dagger}-\beta^{*}\hat{b}\right)}$,
conditioned by the result  $\beta$, on the remote
mode $b$ of the entangled resource yielding the teleported
state. We precise that throughout the paper, the gain of the teleporter is set to unity.
\par
\noindent
Formally, various descriptions of the above protocol
have been presented in the literature, ranging from the Wigner
function formalism originally introduced by Braunstein and
Kimble~\cite{Braunstein}, the Fock state expansion approach~\cite{VanEnk}, the transfer operator description~\cite{Hofmann} to the characteristic function
based formalism~\cite{Marian}. Although this latter is extensively used
when non-Gaussian entangled resources are involved in quantum
teleportation, we adopt the transfer operator description
that is more suited for our work. So as to assess the performances
of quantum teleportation with the non-Gaussian amplified
twin-beam, we establish the expression of the average
fidelity. In the formalism of the transfer operator, the expressions
of the teleportation fidelity, the probability for the homodyne
detection to display the outcome $\beta$ and the average
fidelity are evaluated through the action of a certain transfer
operator on the quantum states involved in the teleportation
process. The quantum state in Bob's possession arising
from the whole teleportation process (transfer of information
through the quantum and classical channels) and that we dub ”output state” is obtained through the action of the transfer operator $\hat{T}(\beta)$ on the input state
\begin{equation}\label{out} 
\left|\psi_{out}(\beta) \right\rangle=\hat{T}(\beta) \left| \psi_{in} \right\rangle,
\end{equation}
we note that the output state under consideration is conditioned
by the result of the homodyne measurement and is thus not normalized. The probability of a given outcome to occur reads
\begin{equation}\label{Ps} 
p(\beta)=\left\langle \psi_{out}(\beta) | \psi_{out}(\beta)\right\rangle.
\end{equation}
Given those expressions, the fidelity of teleportation $\mathcal{F}(\beta)$ defined
as the overlap between the input $\left| \psi_{in} \right\rangle$ and the output  $\left| \psi_{out}(\beta) \right\rangle$ states is found to be
\begin{equation}\label{Fidelity} 
\mathcal{F}(\beta)=\frac{1}{p(\beta)}\left|\left\langle \psi_{in} \right|\hat{T}(\beta) \left| \psi_{in} \right\rangle\right|^{2},
\end{equation}
whereas the average fidelity, being the fidelity of teleportation averaged over the possible outcomes $\beta$ assume the following expression
\begin{equation}\label{Fidelityav} 
\bar{\mathcal{F}}=\int d\beta^{2} p(\beta)\mathcal{F}(\beta)=\int d\beta^{2} \left|\left\langle \psi_{in} \right|\hat{T}(\beta) \left| \psi_{in} \right\rangle\right|^{2}.
\end{equation}
In order to evaluate the quantities of concern, one has to establish
the expression of the transfer operator that depends on the
bipartite entangled resource. In~\cite{Hofmann}, Hofmann \textit{et. al.} draw
up the expression that the transfer operator assumes when the
entangled resource consists of the standard twin-beam
\begin{equation}\label{Transfer} 
\hat{T}(\beta)=\sqrt{\frac{1-\chi^{2}}{\pi}}\sum_{n=0}^{\infty}\chi^{n}\hat{D}(\beta)\ket{n} \bra{n}\hat{D}(-\beta),
\end{equation}
where $\hat{D}(\beta)$ denotes the usual displacement operator acting on
the input state. Moreover, it has been noticed in~\cite{Cochrane} that if
the entangled bipartite state assumes the following form
\begin{equation}\label{Fock} 
\ket{\phi}=\mathcal{N}\sum_{n=0}^{\infty}k_{n}\ket{n,n},
\end{equation}
where $\mathcal{N}$ is a normalisation constant and $k_{n}$ the coefficients
of the entangled resource when expanded in a Fock basis, the transfer operator generalizes to
\begin{equation} 
\hat{T}(\beta)=\frac{\mathcal{N}}{\sqrt{\pi}}\sum_{n=0}^{\infty}k_{n}\hat{D}(\beta)\ket{n} \bra{n}\hat{D}(-\beta),
\end{equation}
As we can clearly see, our entangled state arising from the
action of an optimal NLA on a standard EPR-state is of the
form Eq.~(\ref{Fock}) where the normalisation constant and coefficients respectively read as follows
\begin{equation} 
\mathcal{N}=\sqrt{\frac{1-\chi^{2}}{P_{s,\chi}}},
\end{equation}
\begin{equation}
k_{n}=
\begin{cases}
g^{n-p}\chi^{n}, & \text{if}\ n\leq p \\
\chi^{n}, & \text{otherwise}
\end{cases}
\end{equation}

\subsection{Teleportation of coherent states}

In order to assess the performances of the entangled resource considered in our work, we focus on the paradigmatic case of coherent states teleportation. For the sake of comparison, we first recall some results regarding the standard protocol
with the two-mode squeezed vacuum as entangled state. The output state arising from the teleportation process performed on an input coherent state of amplitude $\alpha$ is found to be
\begin{equation}\label{coherent}
\hat{T}(\beta)\ket{\alpha}=\sqrt{\frac{1-\chi^{2}}{\pi}}\exp{\left[-(1-\chi^{2})\frac{\left| \alpha-\beta \right|^{2}}{2}\right]}\exp{\left[(1-\chi)\frac{ \alpha\beta^{*}-\alpha^{*}\beta}{2}\right]}\ket{\chi(\alpha-\beta)}.
\end{equation}

\begin{figure} [htbp]
	%\vspace*{13pt}
	\centerline{ \epsfig{file=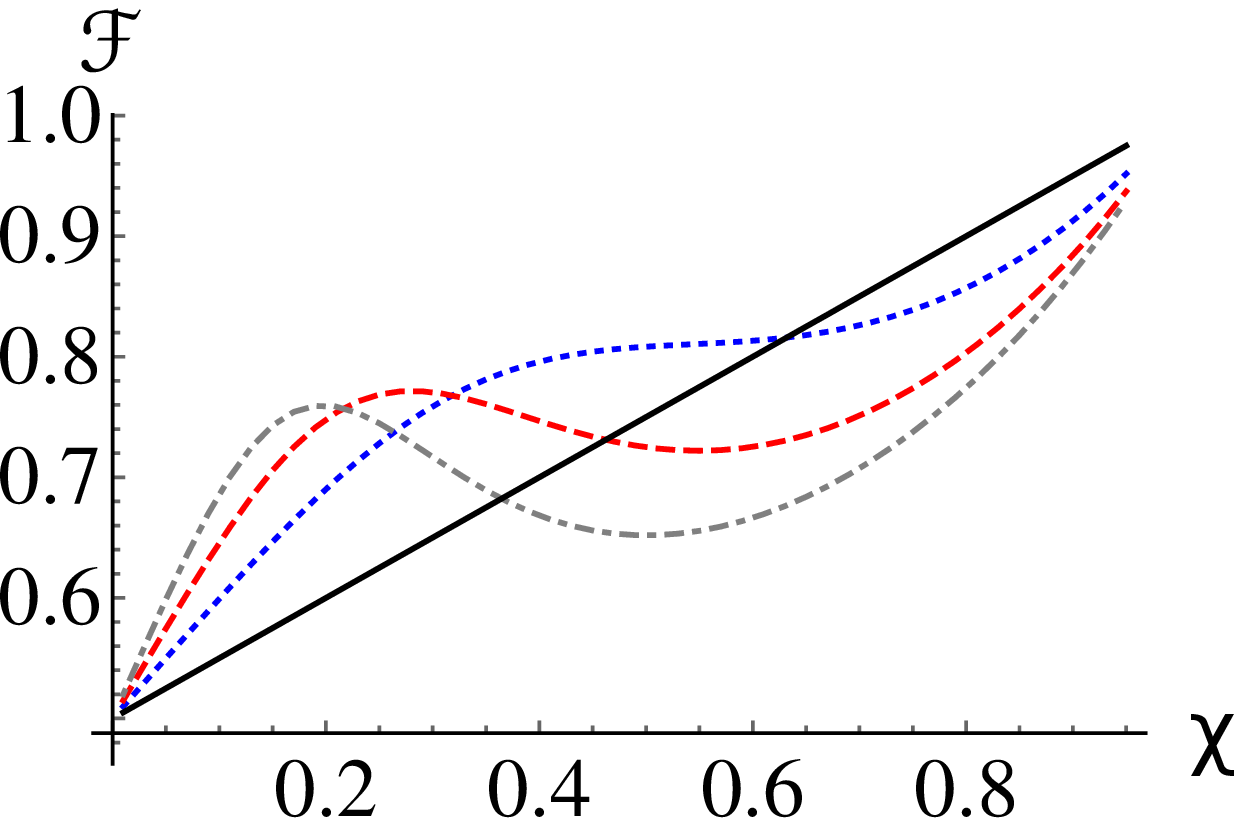,width=6cm}
		\epsfig{file=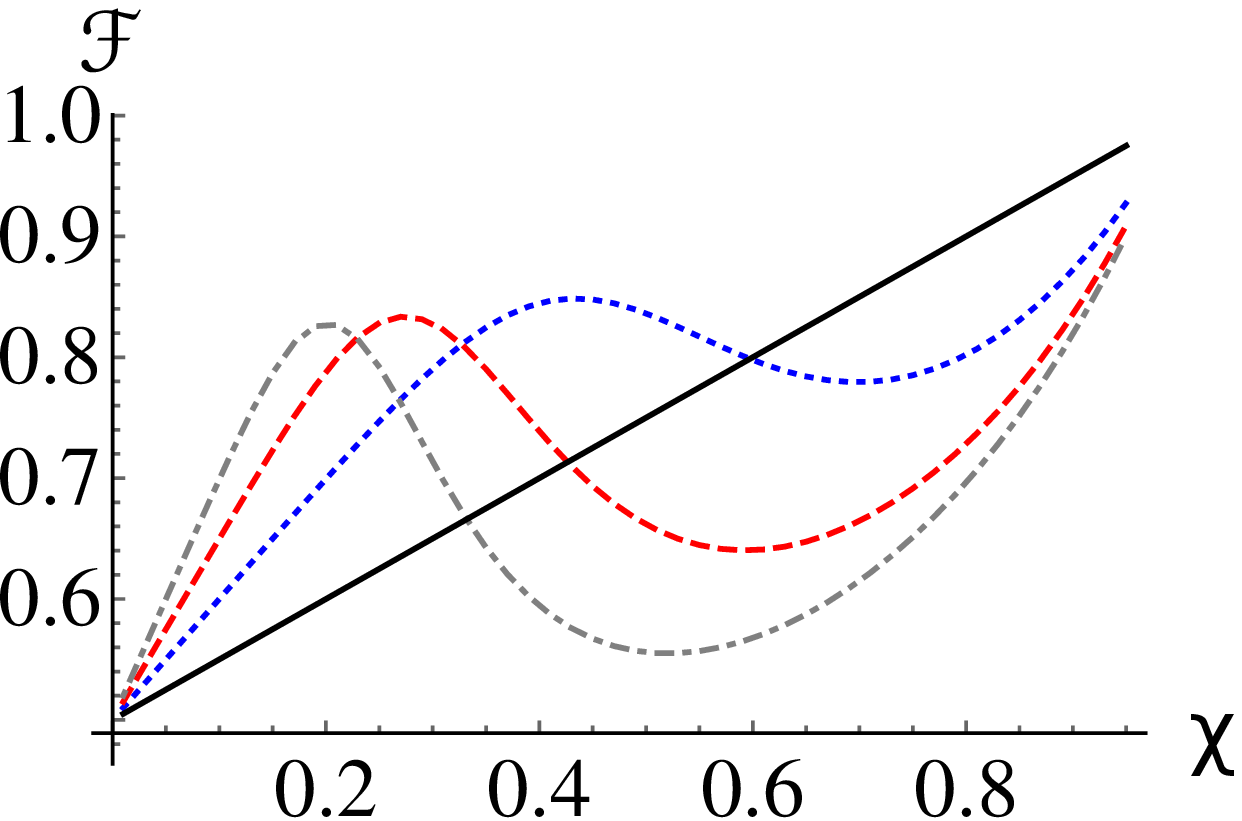,width=6cm}}
	\centerline{\epsfig{file=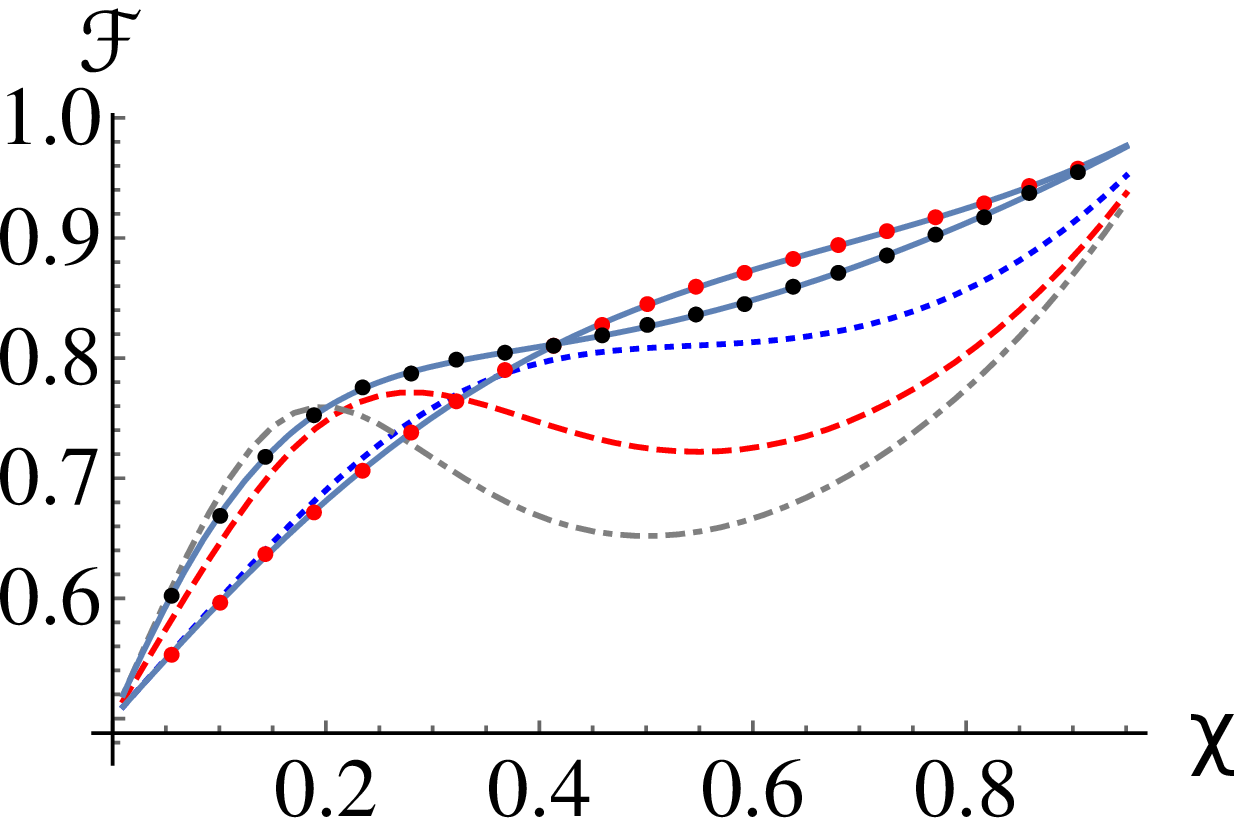,width=6cm}
		\epsfig{file=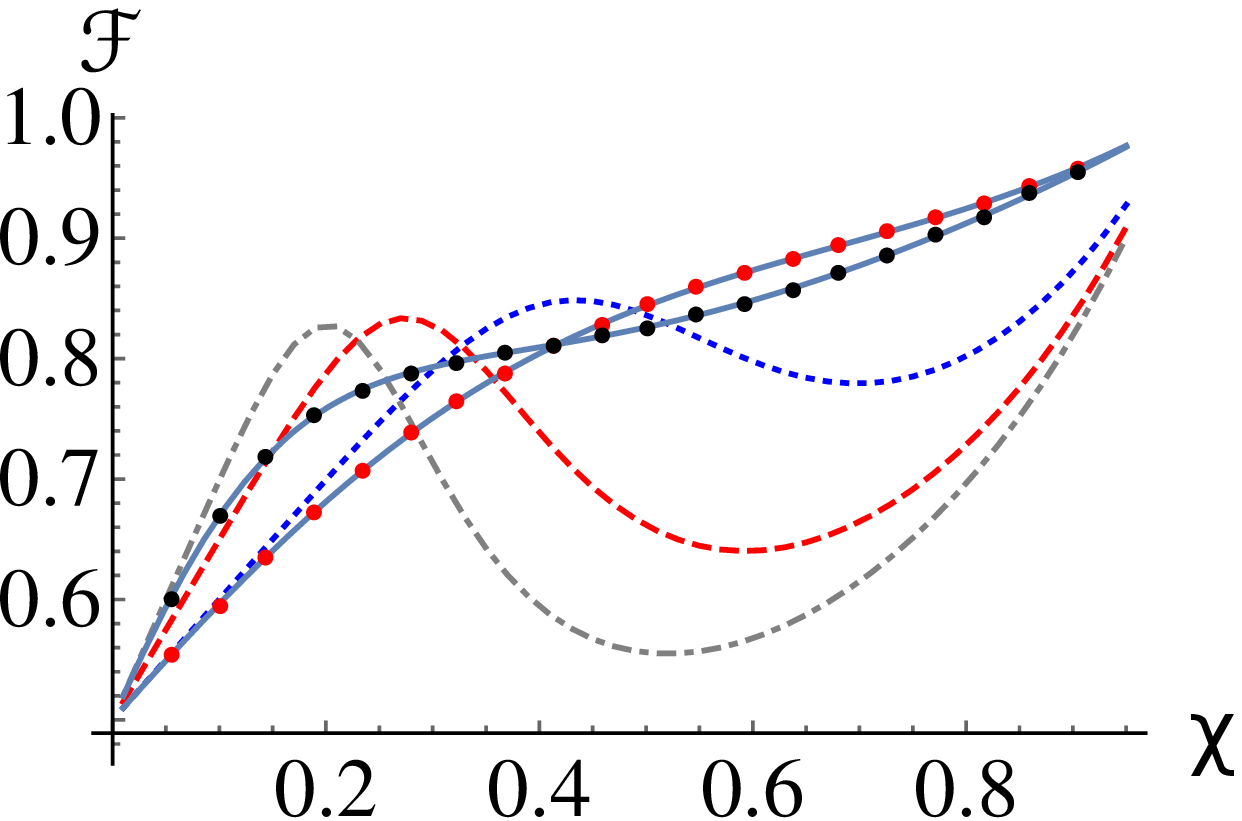,width=6cm}}
	
	\vspace*{13pt}
	\fcaption{(Color online) The average fidelity $\bar{\mathcal{F}}$ as a function of $\chi$ for a two mode squeezed vacuum (black full line), photon-subtracted twin-beam (red circles), photon-added-then-subtracted twin-beam (black circles) and amplified twin-beams generated from different configurations of the NLA: $g=2$ (blue dotted line), $g=3$ (red dashed line), $g=4$ (grey dotted dashed line). In the left panels, p is setted to 2 while in the right panels $p=4$.} \label{f:Fidelity}
\end{figure}

We note that the output is also a coherent state whose amplitude coincides with the displaced then attenuated input amplitude respectively by the factors $\beta$ and $\chi$. The teleportation fidelity averaged over all the homodyne measurement outcomes $\beta$ is independent of the input amplitude and reads
\begin{equation}\label{Fidaverage} 
\bar{\mathcal{F}}=\frac{1}{2}(1-\chi).
\end{equation}
Regarding our B-K like protocols where the standard EPR
state is substituted by the conditional amplified twin-beam,
the fidelity of teleportation and the probability of occurrence
for a given outcome $\beta$ are calculated by means of the following
formula
\begin{equation}
\bra{n}\hat{D}(\beta)\ket{\alpha}=(n!)^{-1/2}(\alpha+\beta)^{n}\exp{
	\frac{1}{2}\left| \alpha+\beta \right|^{2}}\exp{\frac{1}{2} (\alpha^{*}\beta-\alpha\beta^{*})}.
\end{equation}
\begin{figure} [htbp]
	%\vspace*{13pt}
	\centerline{ \epsfig{file=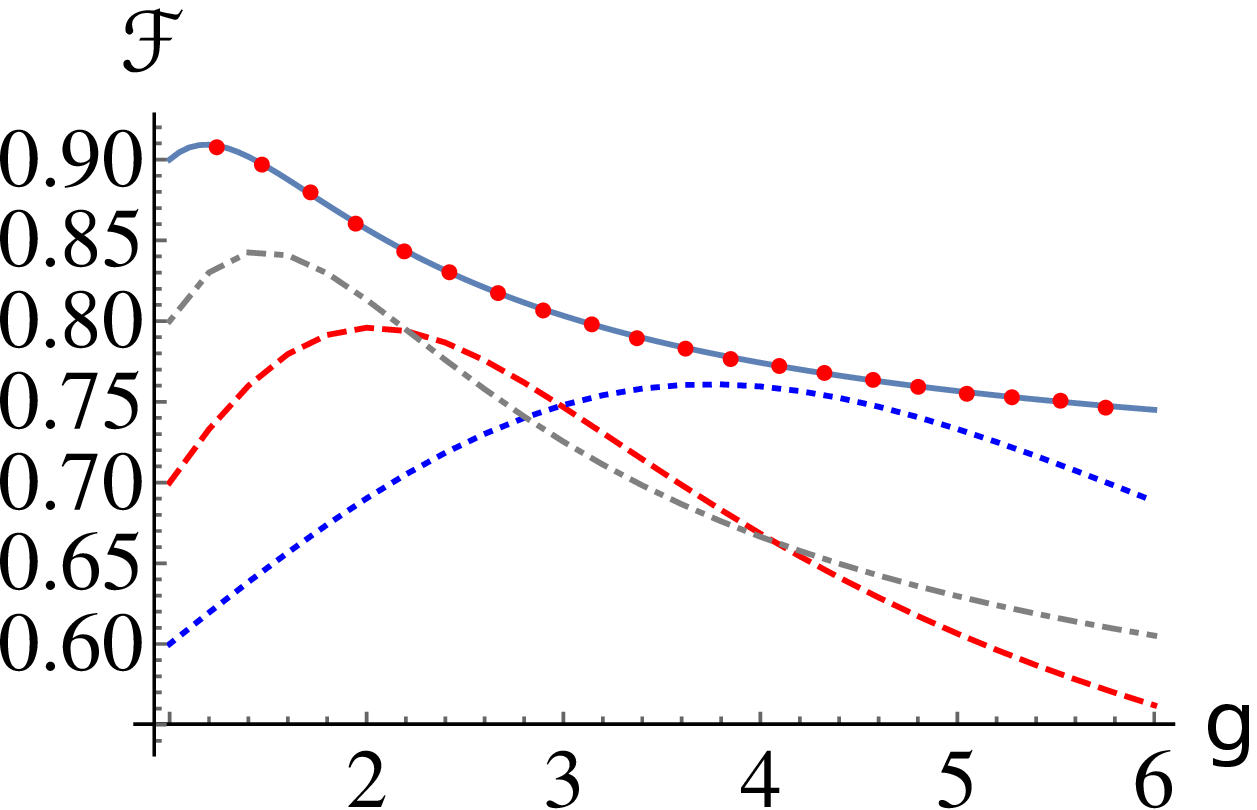,width=6cm}
		\epsfig{file=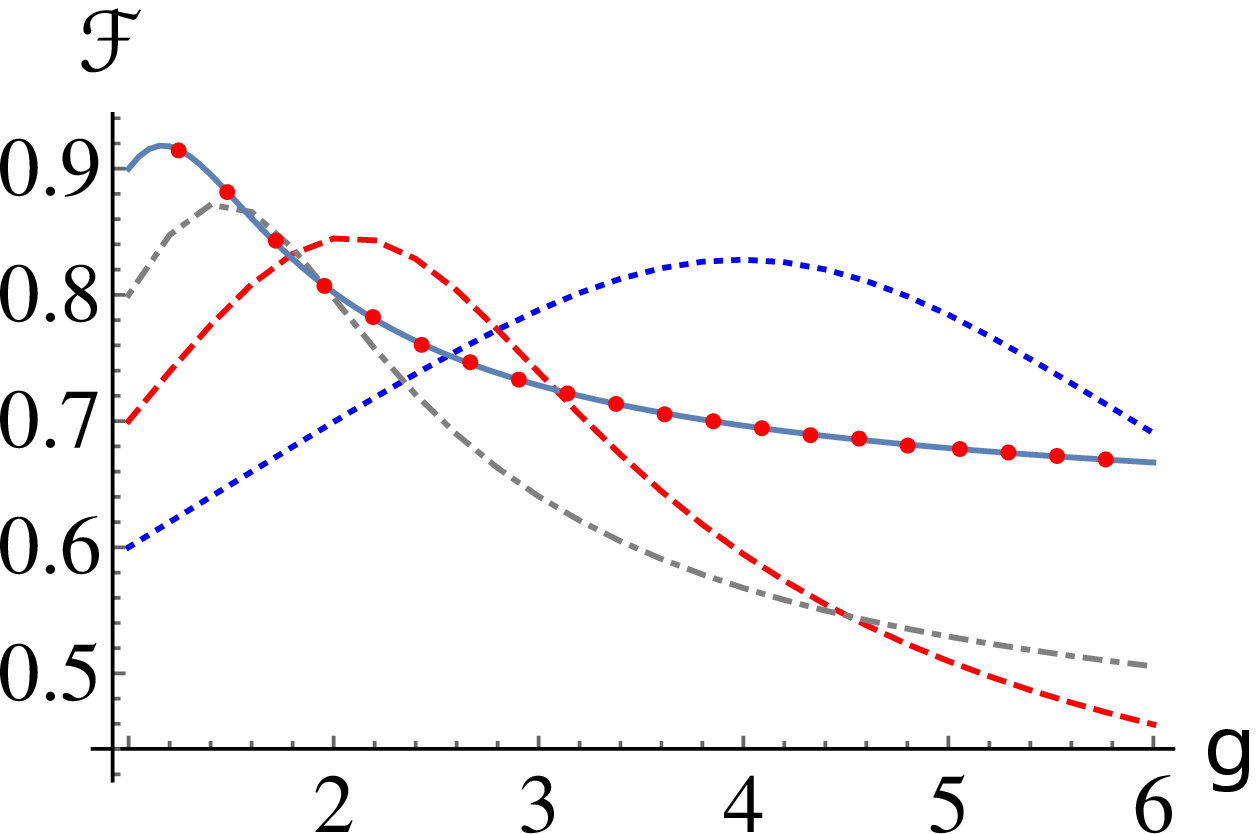,width=6cm}}
	
	\vspace*{13pt}
	\fcaption{(Color online) The average fidelity $\bar{\mathcal{F}}$ as a function of $g$ for entangled resources with different input energies: $\chi=0.22$ (blue dotted line), $\chi=0.4$ (red dashed line), $\chi=0.6$ (grey dotted dashed line) and $\chi=0.8$ (red circles). In the left panel, p is fixed at $2$ whereas in the right panel $p=4$.} \label{f:optimized_fidelity}
\end{figure}
The average fidelity is then carried out numerically by truncating
the Fock basis to a relevant order. In Fig.~\ref{f:Fidelity} are
reported the plots of the average fidelity $\bar{\mathcal{F}}$ for input coherent state $\alpha=2$ as a function of the squeezing parameter $\chi$. A
comparative survey between the non-deterministically amplified twin-beams
and their standard version along with different instances of commonly used de-Gaussified twin-beams is drawn up. As it is apparent from the two sub-figures on top, there is a range of the squeezing parameter where
the average fidelity for quantum teleportation with our conditional entangled resource is enhanced when compared with the standard B-K protocol. We remark that although the amplified twin-beams present a higher excess Von Neumann entropy regardless of the squeezing parameter and the NLA parameters, the average fidelity is enhanced only in a certain region of the input energy. In addition, we notice that
the critical value delimiting that region depends on the NLA gain $g$ for a fixed value of the truncation order $p$ in such a way that the strongest is the amplification the smallest is that critical value. 
Through sub-figures on bottom of Fig.~\ref{f:Fidelity}, we compare the performances of coherent states teleportation protocol exploiting the de-Gaussification process based on an optimal NLA and two well-known de-Gaussification schemes that respectively involve photons subtraction and photons addition then subtraction. Expressions of their average fidelities are independent of the input coherent amplitude and are reported in~\cite{Yang}, where, it was also shown that the photon-added-then-subtracted twin-beam beats the performance of the photon-subtracted twin-beam in the range of weak energies. It appears that, for strong calibrations of the NLA (high values of $g$ and $p=4$), the amplified twin-beams outperform the photon-added-then-subtracted two-mode squeezed state when the input energy belongs to a certain interval, whereas being detrimental to the quality of teleportation in the remaining region. A different behaviour is observed when the threshold is fixed at $p=2$. Indeed, when the gain is below $g=3$, enhancement is achieved compared with the photon-subtracted resource whereas the quality of teleportation remains below that of the photon-added-then-subtracted twin-beam. Finally, We notice that numerical results of the average fidelity for a broad set of input amplitudes ranging from $-20$ to $20$ that includes also complex values show an identical behaviour, we thus state that the teleportation quality captured by the fidelity is state independent. Hence, the plots produced for the specific value ($\alpha=2$) of the input amplitude considered throughout the paper generalize to arbitrary coherent input state.  
\par

As it appears from the upshots displayed in Fig.~\ref{f:Fidelity}, noiseless amplifications operating in the strong working regime (high values of the gain) tend to better enhance the average fidelity when entangled resources with low energies are considered whereas being detrimental for the teleportation quality for EPR states with high squeezing parameter. Identification of the values of the gain leading to the most substantial improvement afforded by the NLA is of great interest. In Fig.~\ref{f:optimized_fidelity}, the average fidelity of teleporting coherent states  with entangled resources of different squeezing parameters is plotted as a function of the NLA's gain. The intersections of the plots with the "y" axis characterize the average fidelity of the standard B-K protocol. As expected, and in accordance with our prior observations, the improvements induced by the NLA on the teleportation quality are optimized by high values of the gain for twin-beams with low input energies while weak amplifications lead to more substantial fidelities when performed on high squeezed entangled resources. Due to the tunable nature of the NLA, its employment maybe optimized with respect to its intrinsic parameters for each considered EPR states and hence take full advantage of the de-Gaussification process. We notice that beyond a certain value of the input squeezing of the twin-beam, the NLA is no more useful irrespectively of its calibration.

An interesting feature of non-deterministically amplified EPR states is thus to lead up high fidelities starting from a standard EPR state of weak energy whose experimental generation is well held down. Hence, so as to achieve high average fidelities for the continuous variable quantum teleportation, one may exploit the advantages of the optimal NLA that acts as an entanglement distiller instead of seeking to engineer intense squeezers. Finally, we point out that even when moderated amplifications are implemented, the amplified entangled resources show, in a certain region of weak energies, a more substantial improvement than that of the photon-subtracted EPR state extensively studied in previous works~\cite{Dell'Anno,Yang,Seshadreesan}. A similar observation stands for its inconclusive version~\cite{Olivares}.

\begin{figure} [htbp]
	%\vspace*{13pt}
	\centerline{\epsfig{file=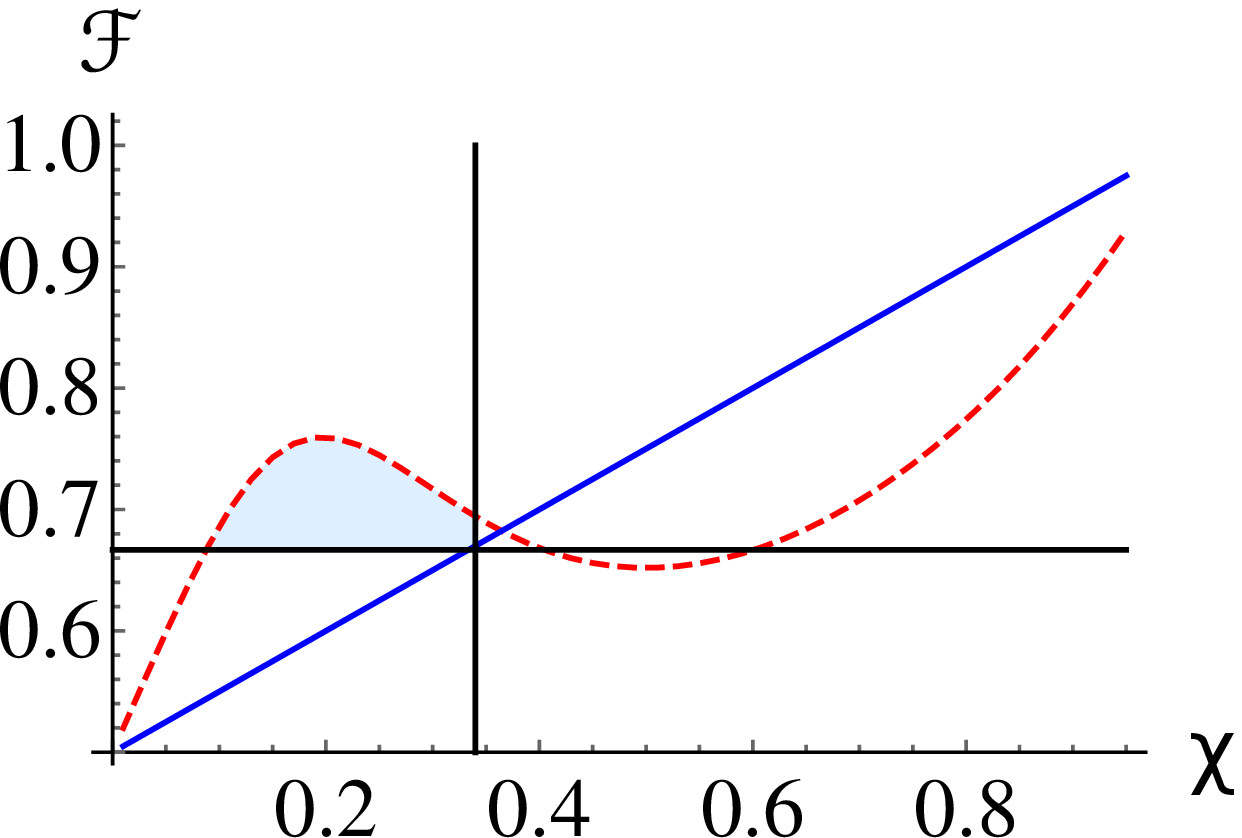, width=8.2cm}} %100 percent
	\vspace*{13pt}
	\fcaption{ (Color online) Dependence of the quality of teleporting a
		coherent state of amplitude $\alpha=2$ with respect to the squeezing
		parameter $\chi$ by means of the standard twin-beam (blue solid line) and
		the amplified twin-beam under the calibration ($g=2$ and $p=4$) (red dotted line). The light-blue shaded area represents where
		the amplified twin-beam achieves an average fidelity exceeding the
		boundary $2/3$ while the standard one does not. The horizontal and
		vertical lines delimit the region of concern.}\label{fig:dashedarea}
\end{figure}

\bigskip
Information transmission in a teleportation process is said truly nonlocal if its average fidelity $\bar{\mathcal{F}}$ exceeds $1/2$~\cite{Braunstein2}. As shown in Fig.~\ref{f:Fidelity}, both the standard and amplified twin-beams surpass this limit independently of the NLA configuration. A more significant boundary ($\bar{\mathcal{F}}>2/3$), witnessing that the state in Bob's possession is the best existing copy was derived in~\cite{Ralph}. The meaning of this boundary
is that when a secure transmission is required, even if Bob is not able to avert an eavesdropping action that alters or clones the transferred state, he can check if he's the only owner of this latter~\cite{Cerf,Grosshans}. Indeed, this is made possible by considering the average fidelity of the teleportation process. When
the fidelity beats the boundary $2/3$, Bob knows unambiguously that his state was not duplicated and the teleportation is said ”secure”. In Fig.~\ref{fig:dashedarea} are plotted the average fidelities of the standard twin-beam and its non-deterministically amplified homologous with the configuration ($g=2$ and $p=4$) as functions of the squeezing parameter $\chi$, and where a coherent state of amplitude $\alpha=2$ is teleported. Fig.~\ref{fig:dashedarea} highlights a region (the light-blue shaded area) where the average fidelity of the teleportation with our non-Gaussian resource exceeds $2/3$ whereas the quality of the standard protocol with the two-mode squeezed vacuum remains below that boundary. The vertical and horizontal lines serve to mark out the shaded region. According to those results, our proposed conditional scheme based on non-deterministic noiseless amplification enables to improve the quality of coherent state teleportation and makes the process
secure in a region where the standard EPR resource does not.

\subsection{Identification of the source of improvement: entanglement content and non-Gaussianity}

In this section, we will discuss the properties of the amplified
twin-beam that led to enhance the quality of quantum
teleportation. We confront our results on the degree of entanglement
and the non-Gaussianity of our conditional entangled
resource presented in subsesction (\ref{amplifiedTB}) to the content of Fig.~\ref{f:Fidelity} regarding the average fidelity of coherent state teleportation.
First, we focus on the entanglement content quantified by the
excess Von Neumann entropy for different configurations of
the NLA. We observe from Fig.~\ref{f:ExcessVN} and Fig.~\ref{f:Fidelity} that the degree of entanglement for all the considered configurations of the
NLA is higher than that of the standard twin-beam, at the opposite
of the average fidelity, which is degraded whenever the
squeezing parameter exceeds a certain threshold. Thereby, the
excess Von Neumann entropy does not provide full explanation
about the behaviour of the quality of teleportation. Contrariwise,
regarding the non-Gaussian states arising from different
configurations, the more substantial is the excess Von
Neumann entropy, the higher is the average fidelity of the entangled
resource. Hence, we identify the entanglement content
captured by the excess Von Neumann entropy as a relevant
property to compare the quality of teleportation attainable with
the different non-Gaussian resources, though not sufficient to elucidate its whole behaviour. 

\bigskip
Naturally, the next step is to examine how the amount of
non-Gaussianity of the entangled bipartite states
generated from the NLA action acts on the quality of the teleportation.
For that purpose, we jointly analyse the results of
Figs.~\ref{fig:non-Gaussianity} and Fig.~\ref{f:Fidelity}. It is clear that there is no trivial dependence between the entropic non-Gaussianity and the fidelity of teleportation for all the considered configurations. Moreover, in
the range of high values of the squeezing parameter, the 
non-Gaussianity seems to be detrimental to the quality of teleportation.
Entropic non-Gaussianity doesn't offer any clarification
on the origin of the improvement observed for the
teleportation quality.
\bigskip

At last, we consider the dependence of the average fidelity
of the quantum teleportation with respect to the EPR correlation
that characterizes the entanglement content of the bipartite
states. Based on the results reported in Fig.~\ref{f:EPR} and Fig.~\ref{f:Fidelity},
we establish a comparison between the two quantities of interest.
First of all, we notice two distinct regions for both the
EPR correlation and the quality of the teleportation for a given
configuration of the amplified twin-beam. For $0<\chi<\chi_{c1}$ ,
where $\chi_{c1}$ is a critical value that depends on the NLA's calibration,
the entanglement content captured by the EPR correlation
is lowered when compared with that of the standard
EPR state, whereas in the range of squeezing parameters surpassing
$\chi_{c1}$ , the EPR correlation is greater than that of the
standard twin-beam. A similar behaviour is observed for the
average fidelity where the two regions are delimited by $\chi_{c2}$ 
instead of $\chi_{c1}$. Furthermore, in the range of weak energies,
it appears that the more intense is the amplification, the more
correlated are the two-modes and better is the quality of teleportation.
Our results suggest that EPR correlation is a good witness for an improved average fidelity of the teleportation and may explain its origin. Finally, we remark that in the regime of intense amplifications ($g=3,4$ and $p=4$) the presence of an area where $\Delta z^{2}$ indicates non-correlated resource
and still leads to an average fidelity $\bar{\mathcal{F}}>1/2$. This observation
meets the result highlighted in~\cite{Lee,Wang} according to
which EPR correlation is not a necessary condition for quantum
teleportation. 

\section{Conclusion}\label{s:sec4}
In summary, we have examined the properties of the non-Gaussian bipartite 
states resulting from the non-deterministic
noiseless linear amplification of two-mode
squeezed vacua. Various characteristic attributes related to
the entanglement content and the non-Gaussian character of
the amplified non-Gaussian resources have been considered
and then compared with respect to the different configurations
of the NLA and those of the standard twin-beam. We show
that the excess Von Neumann entropy of the generated states
is greater than that of the standard twin-beam independently
of the configuration of the NLA. In particular, weak amplifications
promote the increase of the degree of entanglement in
the range of strong squeezing whereas intense calibrations of the NLA
tend to optimise it in the interval of weak energies. Regarding
the EPR correlation, a different behaviour has been observed:
the amplification appears to enhance the entanglement content
in the region of low energies whereas being detrimental when the squeezing of the input standard twin-beam exceeds a certain critical value. 
\par
In the light of these results, we have investigated a B-K like teleportation 
protocol where the standard entangled resource is substituted by the non-deterministically amplified twin-beam. We then drew up
an elaborated comparison of the performance of the amplified non-Gaussian entangled resource and its Gaussian homologous in the quantum teleportation of coherent input light. Our upshot identify two range of the input energy delimited by a critical value that depends on the NLA calibration: for an input energy lower
than the critical value, substantial improvement of the quality
of teleportation can be reached whereas in the remaining region,
amplification appears to be detrimental to the average fidelity.
Thereby, the NLA is identified as a robust resource for
quantum teleportation when performed on initial EPR-states
with low energies. Furthermore, we emphasize the existence
of a region where the non-Gaussian resources achieve secure
quantum teleportation while the standard Gaussian bipartite
state does not. Finally, we show that the enhancement of the
quality of teleportation and the EPR correlation pursue approximately
the same behaviour in opposition with the excess
Von Neumann entropy and the non-Gaussianity that show a
non-trivial dependance. Thus, we identify EPR correlation
as a good indicator to gauge the quality of teleportation that
may explain the origin of the noticeable improvement.

\nonumsection{Acknowledgements}
\noindent The authors are grateful to Hakim Gharbi, Marco Genoni, Francesco Albarelli and 
Matteo Bina for support and useful discussions.

\nonumsection{References}


\begin{thebibliography}{000}
    \bibitem{Nielsen}
    M. A. Nielsen and I. L. Chuang (2010), {\it Quantum Computation and
		Quantum Information}, University Press (Cambridge).
	
	\bibitem{Bennett}
	C. H. Bennett, G. Brassard, C. Cr\'epeau, R. Jozsa, A. Peres, and
	W. K. Wootters (1993), {\it Teleporting an unknown quantum state via dual classical and Einstein-Podolsky-Rosen channels},
	Physical review letters, Vol. 70, No 13, p. 1895.
	
	\bibitem{Braunstein}
	S. L. Braunstein and H. J. Kimble (1998), {\it Teleportation of Continuous Quantum Variables},
	Physical review letters, Vol. 80, No 4, p. 869.
	
	\bibitem{Furusawa} 
	A. Furusawa, J. L. Sørensen, S. L. Braunstein, C. A. Fuchs,
	H. J. Kimble, and E. S. Polzik (1998), {\it Unconditional Quantum Teleportation} Science, Vol. 282, No. 5389, pp. 706-709.
	
	
	\bibitem{Yukawa} 
	M. Yukawa, H. Benichi, and A. Furusawa (2008), {\it High-fidelity continuous-variable quantum teleportation toward multistep quantum operations},
	Physical Review A, Vol. 77, No. 2, p. 022314.
	\bibitem{Eisert}
	J. Eisert, S. Scheel, and M. B. Plenio (2002), {\it Distilling Gaussian States with Gaussian Operations is Impossible}, Physical review letters, Vol. 89, No. 13, p. 137903.
	
	\bibitem{Fiurasek}
	J. Fiur\'a\ifmmode \check{s}\else \v{s}\fi{}ek (2002), {\it Gaussian Transformations and Distillation of Entangled Gaussian States}, Physical review letters, Vol. 89, No. 13, p. 137904.
	
	\bibitem{Niset}
	J. Niset, J. Fiur\'a\ifmmode \check{s}\else \v{s}\fi{}ek, and N. J. Cerf (2009),{\it No-Go Theorem for Gaussian Quantum Error Correction}, Physical review letters, Vol. 102, No. 12, p. 120501.
	
	\bibitem{Eberle}
	T. Eberle, V. H\"{a}ndchen, and R. Schnabel (2013),{\it Stable control of 10 dB two-mode squeezed vacuum states of light}, Optics express, Vol. 21, No. 9, pp. 11546-11553.
	
	\bibitem{Vahlbruch}
	
	H. Vahlbruch, M. Mehmet, K. Danzmann, and R. Schnabel (2016), {\it Detection of 15 dB squeezed states of light and their application for the absolute calibration of photoelectric quantum efficiency}, Physical review letters, Vol. 117, No. 11, p. 110801.
	
	\bibitem{Opatrny}
	T. Opatrn\'y, G. Kurizki, and D.-G. Welsch (2000), {\it Improvement on teleportation of continuous variables by photon subtraction via conditional measurement}, Physical Review A, Vol. 61, No. 3, p. 032302.
	
	\bibitem{Cochrane}
	P. T. Cochrane, T. C. Ralph, and G. J. Milburn (2002), {\it Teleportation improvement by conditional measurements on the two-mode squeezed vacuum}, Physical Review A, Vol. 65, No. 6, p. 062306.
	
	\bibitem{Dell'Anno}
	F. Dell’Anno, S. De Siena, L. Albano, and F. Illuminati (2007), {\it Continuous-variable quantum teleportation with non-Gaussian resources} Physical Review A, Vol. 76, No. 2, p. 022301.
	
	\bibitem{Yang}
	Y. Yang and F.-L. Li (2009), {\it Entanglement properties of non-Gaussian resources generated via photon subtraction and addition and continuous-variable quantum-teleportation improvement}, Physical Review A, Vol. 80, No. 2, p. 022315.
	
	\bibitem{Olivares}
	S. Olivares, M. G. A. Paris, and R. Bonifacio (2003), {\it Teleportation improvement by inconclusive photon subtraction}, Physical Review A, Vol. 67, No. 3, p. 032314.
	
	\bibitem{Olivares2}
	S. Olivares and M. G. A. Paris (2004), {\it Enhancement of nonlocality in phase space} Physical Review A, Vol. 70, No. 3, p. 032112.
	
	\bibitem{Adnane}
	H. Adnane, M. Bina, F. Albarelli, A. Gharbi, and M. G. Paris (2019), { \it Quantum state engineering by non-deterministic noiseless linear amplification}, Physical Review A, Vol. 99, No. 6, p.  063823. 
	
	\bibitem{Menzies}
	D. Menzies and S. Croke (2009), {\it Noiseless linear amplification via weak measurements}, quant-ph/0903.4181.
	
	\bibitem{McMahon}
	N. A. McMahon, A. P. Lund, and T. C. Ralph (2014), {\it Optimal architecture for a nondeterministic noiseless linear amplifier}, Physical Review A, Vol. 89, No. 2, p. 023846.
	
	\bibitem{Pandey}
	
	S. Pandey, Z. Jiang, J. Combes, and C.M.Caves (2013), {\it Quantum limits on probabilistic amplifiers}, Physical. Review. A, Vol.88, No.3, p. 033852.
	
	\bibitem{Barnett}
	S. M. Barnett and S. J. D. Phoenix (1989), {\it Entropy as a measure of quantum optical correlation}, Physical Review A, Vol. 40, No. 5, p. 2404.
	
	\bibitem{Vedral}
	V. Vedral, M. B. Plenio, K. Jacobs, and P. L. Knight (1997), {\it Statistical inference, distinguishability of quantum states, and quantum entanglement}, Physical Review A, Vol. 56, No. 6, p. 4452.
	
	\bibitem{Genoni}
	M. G. Genoni, M. G. A. Paris, and K. Banaszek (2008), {\it Quantifying the non-Gaussian character of a quantum state by quantum relative entropy}, Physical Review A, Vol. 78, No. 6, p. 060303.
	
	\bibitem{Ralph2}
	
	T. C. Ralph (2011), {\it Quantum error correction of continuous-variable states against Gaussian noise}, Physical Review A, Vol. 84, No 2, p. 022339.
	
	\bibitem{Loudon} 
	R. Loudon (2000), {\it The quantum theory of light}, OUP (Oxford).
	
	\bibitem{Araki}
	H. Araki and E. H. Lieb (1970), {\it Entropy inequalities}, Communications in Mathematical Physics, Vol. 18, No. 2, pp 160-170.  
	
	\bibitem{Einstein}
	A. Einstein, B. Podolsky, and N. Rosen (1935), {\it Can Quantum-Mechanical Description of Physical Reality Be Considered Complete?}, Physical review, Vol. 47, No. 10, p. 777.
	
	\bibitem{Duan}
	L.-M. Duan, G. Giedke, J. I. Cirac, and P. Zoller (2000),  {\it Inseparability Criterion for Continuous Variable Systems}, Physical Review Letters, Vol. 84, No. 12, p. 2722.
	
	\bibitem{Popescu}
	S. Popescu and D. Rohrlich (1997), {\it Thermodynamics and the measure of entanglement}, Physical Review A, Vol. 56, No. 5, p. R3319.
	
	\bibitem{Ralph1}
	
	T. C. Ralph and A. P. Lund (2009), {\it Nondeterministic noiseless linear amplification of quantum systems}, AIP Conference Proceedings. Vol. 1110. No. 1. AIP.
	
	\bibitem{Xiang}
	
	G. Y. Xiang, T. C. Ralph, A. P. Lund, N. Walk, and G. J. Pryde (2010),{\it Heralded noiseless linear amplification and distillation of entanglement}, Nature Photonics, vol. 4, No. 5, p. 316.
	
	\bibitem{Ferreyrol}
	
	F. Ferreyrol, M. Barbieri, R. Blandino, S. Fossier, R. Tualle-Brouri, and P. Grangier (2010), { \it Implementation of a nondeterministic optical noiseless amplifier}, Physical review letters Vol. 104, No. 12, p. 123603.
	
	\bibitem{Ferreyrol1}
	
	F. Ferreyrol, R. Blandino, M. Barbieri, R. Tualle-Brouri, and P. Grangier (2011), {\it Experimental realization of a nondeterministic optical noiseless amplifier}, Physical Review A, Vol. 83, No. 6, P. 063801.
	
	\bibitem{Usaga}
	
	M. A. Usuga, C. R. Muller, C. Wittmann, P. Marek, R. Filip, C. Marquardt, G. Leuchs, and U. L. Andersen (2010),{\it Noise-powered probabilistic concentration of phase information}, Nature Physics, Vol. 6, No. 10, p. 767.
	
	\bibitem{Jarumir}
	
	J. Fiur\'a\ifmmode \check{s}\else \v{s}\fi{}ek (2009), {\it Engineering quantum operations on travelling light beams by multiple photon addition and subtraction}, Physical Review A, Vol. 80, No. 5, p. 053822.
	
	\bibitem{Zavatta}
	
	A. Zavatta, J. Fiur\'a\ifmmode \check{s}\else \v{s}\fi{}ek and M. Bellini (2011), {\it A high-fidelity noiseless amplifier for quantum light state}, Nature Photonics Vol. 5, No. 1, p. 52.
	
	\bibitem{Jarumir1}
	
	J. Fiur\'a\ifmmode \check{s}\else \v{s}\fi{}ek and N. J. Cerf (2012), {\it Gaussian postselection and virtual noiseless amplification in continuous-variable quantum key distribution}, Physical Review A, Vol. 86, No. 6, p. 060302.
	
	\bibitem{Walk}
	
	N. Walk, T.C. Ralph, T. Symul, and P. K. Lam (2013), {\it Security of continuous-variable quantum cryptography with Gaussian postselection}, Physical Review A, Vol. 87, No. 2, p. 020303.
	
	\bibitem{Chrzanowski}
	
	H. M. Chrzanowski, N. Walk, S. M. Assad, J. Janousek, S. Hosseini, T. C. Ralph, T. Symul, and P. K. Lam (2014), {\it Measurement-based noiseless linear amplification for quantum communication}, Nature Photonics, Vol. 8, No. 4, p. 333.
	
	
	\bibitem{Haw}
	
	J. Y. Haw, J. Zhao, J. Dias, S. M. Assad, M. Bradshaw, R. Blandino, T. Symul, T. C. Ralph, and P. K. Lam (2016), {\it Surpassing the no-cloning limit
		with a heralded hybrid linear amplifier for coherent states}, Nature Communications, Vol.7, p. 13222.
	
	
	\bibitem{Bose}
	S. Bose and M. Kumar (2018), {\it Quantum Teleportation with a Class of Non-Gaussian Entangled Resources}, quant-ph/1804.00190.
	
	\bibitem{Vaidman2}
	L. Vaidman (1994), {\it Teleportation of quantum states}, Physical Review A, Vol. 49, No. 2, p. 1473.
	
	\bibitem{VanEnk}
	S. J. van Enk (1999), {\it Discrete formulation of teleportation of continuous variables}, Physical Review A, Vol. 60, No 6, p. 5095.
	
	\bibitem{Hofmann}
	H. F. Hofmann, T. Ide, T. Kobayashi, and A. Furusawa (2000), {\it Fidelity and information in the quantum teleportation of continuous variables}, Physical Review A, Vol. 62, No. 6, p. 062304.
	
	\bibitem{Marian}
	P. Marian and T. A. Marian (2006), {\it Continuous-variable teleportation in the characteristic-function description}, Physical Review A, Vol. 74, No. 4, p. 042306.
	
	\bibitem{Seshadreesan}
	K. P. Seshadreesan, J. P. Dowling, and G. S. Agarwal (2015), {\it Non-Gaussian entangled states and quantum teleportation of Schrödinger-cat states}, Physica Scripta, Vol. 90, No. 7, p. 074029.
	
	\bibitem{Braunstein2}
	S. L. Braunstein, C. A. Fuchs, and H. J. Kimble (2000), {\it Criteria for continuous-variable quantum teleportation}, Journal of Modern Optics, Vol. 47, No. 2-3, p. 267-278.
	
	\bibitem{Ralph}
	T. C. Ralph and P. K. Lam (1998), {\it Teleportation with Bright Squeezed Light},
	Physical review letters, Vol. 81, No. 25, p. 5668.
	
	\bibitem{Cerf}
	N. J. Cerf, A. Ipe, and X. Rottenberg (2000), {\it Cloning of Continuous Quantum Variables}, Physical Review Letters, Vol. 85, No. 8, p. 1754.
	
	\bibitem{Grosshans}
	F. Grosshans and P. Grangier (2001), {\it Quantum cloning and teleportation criteria for continuous quantum variables}, Physical Review A, Vol. 64, No. 1, p. 010301.
	
	\bibitem{Lee}
	
	S.-Y. Lee, S.-W. Ji, H.-J. Kim, and H. Nha (2011), {\it Enhancing quantum entanglement for continuous variables by a coherent superposition of photon subtraction and addition}, Physical Review A, Vol. 84, No. 1, p. 012302.
	
	\bibitem{Wang}
	S. Wang, L.-L. Hou, X.-F. Chen, and X.-F. Xu (2015), Physical Review A, Vol. 91, No. 6, p. 063832.
	

\end{thebibliography}
\end{document}